\documentclass[onecolumn,showpacs,preprintnumbers]{revtex4}
\usepackage{amssymb}
\usepackage{amsfonts}
\usepackage{amsmath}
\usepackage{graphicx}
\usepackage{dcolumn}
\usepackage{bm}
\usepackage{epsfig}

\begin{document}

\title{Manifest Covariant Hamiltonian Theory of General Relativity}
\author{Claudio Cremaschini}
\affiliation{Institute of Physics, Faculty of Philosophy and Science, Silesian University
in Opava, Bezru\v{c}ovo n\'{a}m.13, CZ-74601 Opava, Czech Republic}
\author{Massimo Tessarotto}
\affiliation{Department of Mathematics and Geosciences, University of Trieste, Via
Valerio 12, 34127 Trieste, Italy\\
Institute of Physics, Faculty of Philosophy and Science, Silesian University
in Opava, Bezru\v{c}ovo n\'{a}m.13, CZ-74601 Opava, Czech Republic}
\date{\today }

\begin{abstract}
The problem of formulating a manifest covariant Hamiltonian theory of
General Relativity in the presence of source fields is addressed, by
extending the so-called \textquotedblleft DeDonder-Weyl\textquotedblright\
formalism to the treatment of classical fields in curved space-time. The
theory is based on a synchronous variational principle for the Einstein
equation, formulated in terms of superabundant variables. The technique
permits one to determine the continuum covariant Hamiltonian structure
associated with the Einstein equation. The corresponding continuum Poisson
bracket representation is also determined. The theory relies on
first-principles, in the sense that the conclusions are reached in the
framework of a non-perturbative covariant approach, which allows one to
preserve both the 4-scalar nature of Lagrangian and Hamiltonian densities as
well as the gauge invariance property of the theory.
\end{abstract}

\maketitle

\section{Introduction}

The Hamiltonian description of classical mechanics, both for discrete and
continuum systems, is of foremost importance and a mandatory prerequisite
for the construction of quantum field theory \cite{Goldstein,dirac5}. The
issue pertains necessarily also General Relativity (GR), and specifically
the so-called Standard Formulation to General Relativity (SF-GR) \cite%
{ein1,hilb,gravi,kuchar}, i.e., Einstein's original approach to his namesake
field equation. The primary goal of this paper is to carry out the
construction of a Hamiltonian description of GR in the context of SF-GR and
such that it leaves unchanged the tensorial form of the Einstein field
equations.\ This work is based on earlier work related to the establishment
of a fully-covariant Lagrangian treatment of SF-GR (\cite{syn}). The latter
is based in turn on the adoption of a new type of variational principle for
continuous fields, i.e. the so-called \emph{synchronous Lagrangian
variational principle}, first introduced in Ref.\cite{syn} (please notice
that below we shall refer to the same paper for all notations which are not
explicitly indicated here). As we intend to show, this choice is also
crucial in the present paper for the establishment of the axiomatic
covariant Hamiltonian treatment.

The task indicated above, however, is by no means straightforward. Indeed,
an axiomatic approach of this type should satisfy precise requisites.\ These
include, in particular, both the Einstein's general covariance principle and
the principle of manifest covariance. It must be stressed that a Hamiltonian
theory of GR fulfilling these requirements is still missing to date.

More precisely, the first requisite states that in the context of GR the
Einstein field equation as well as its solution represented by the metric
tensor $g(r)\equiv \left\{ g_{\mu \nu }(r)\right\} $ should be endowed with
tensor transformation laws with respect to the group of transformations
connecting arbitrary GR-reference frames, i.e., $4-$dimensional curvilinear
coordinate systems spanning the same prescribed space-time $D^{4}\equiv
\left( \mathbf{Q}^{4},g(r)\right) .$\textbf{\ }Here the notation $r\equiv
\left\{ r^{\mu }\right\} ,$ to be used whenever necessary throughout the
paper, actually identifies one of such possible coordinate
parametrizations.\ In SF-GR such a setting is usually identified with the
group of invertible\ local point transformations (i.e., suitably-smooth
diffeomorphism).

Instead, the principle of manifest covariance is actually a particular
realization of the general covariance principle \cite{ein1}. It states that
it should always be possible in the context of any relativistic theory,
i.e., in particular SF-GR, to identify with 4-tensor quantities all
continuum fields, such as the corresponding related Lagrangian and
Hamiltonian continuum variables, as well as the relevant variational and/or
extremal equations involved in the theory. The latter viewpoint is in
agreement with the axiomatic construction originally developed by Einstein
in which both requirements are, in fact, explicitly adopted. Indeed,
according to the Einstein's viewpoint in this reference, in order that
physical laws have an objective physical character, they should not\textbf{\
}depend on the choice of the GR-reference frame. This requisite can only be
met when all classical physical observables and corresponding physical laws
and the mathematical relationships holding among them, are expressed in
tensorial form with respect to the group of transformations connecting the
said GR-frames.

The conjecture that a manifestly covariant Hamiltonian formulation, i.e., a
theory satisfying simultaneously both of these principles, must be possible
for continuum systems is also suggested by the analogous theory holding for
discrete classical particle systems. Indeed, its validity is fundamentally
implied by\ the state-of the-art theory of classical $N-$body systems
subject to non-local electromagnetic (EM) interactions. The issue is
exemplified by the Hamiltonian structure of the EM radiation-reaction
problem in the case of classical extended particles as well as $N-$body EM
interactions among particles of this type \cite%
{EPJ1,EPJ2,EPJ3,EPJ4,EPJ5,EPJ6,EPJ7}, together with the corresponding
non-local quantum theory \cite{EPJ8}.

On the other hand, it should be mentioned that in the case of continuum
fields, the appropriate formalism is actually well-established, being
provided by the DeDonder-Weyl Lagrangian and Hamiltonian treatments \cite%
{donder,weyl,sym1,sym2,sym3,sym4,sym5,sym6,sym7,sym8,sym9}. Such an approach
is originally formulated for fields defined on the Minkowski space-time,
while its extension to classical fields defined in curved space-time is
well-known. Nevertheless, the inclusion of the gravitational field in this
treatment, consistent with the full validity of the manifest covariance
principle, is still missing. The need to adopt an analogous approach also in
the context of classical GR, and in particular for the Einstein equation
itself or its possible modifications, has been recognized before \cite%
{esposito1995,rovelli2002,rovelli2003}. Notice that this feature is of
primary interest in particular in the case of non-perturbative classical and
quantum approaches. In fact, it is known that quantum theories based on
perturbative classical treatments\textbf{\ }are themselves intrinsically
inadequate to establish a consistent theory of quantum gravity.

However, as far as the gravitational field is concerned, a common difficulty
met by all previous non-perturbative Hamiltonian approaches of this type is
that, from the start, they are based on the introduction of Lagrangian
densities which have a non-covariant character, i.e., they are not
4-scalars. A further possible deficiency lies in the choice made by some
authors of non-tensorial Lagrangian coordinates and/or momenta. Such a
concept for example is intrinsic in the Dirac constrained dynamics approach
\cite{cast1,cast2,cast3}. Both features actually prevent the possibility of
establishing a theory in which the canonical variables are tensorial and the
Euler-Lagrange equations, when represented in terms of them, are manifestly
covariant. A possible way-out, first pointed out by De Witt \cite{dewitt67},
might appear that based on the adoption of functional-derivative equations
in implicit form, replacing the explicit Euler-Lagrange equations. This can
be formally achieved provided the variations of the Lagrangian coordinates
and momenta have a tensorial character \cite{dewitt67,rovelli2003}.
Nevertheless, the issue of the possible violation of manifest covariance
remains in place for the definition of the Lagrangian and Hamiltonian
variables.

It must be stressed that a common underlying difficulty still characterizing
these approaches is that the variational Lagrangian densities adopted there
still lack, in a strict sense, the fundamental requirement needed for the
application of the DeDonder-Weyl approach, namely their 4-scalar property.
The physical consequence is that the action functional and consequently also
the variational Lagrangian density do not exhibit the required gauge
properties, which apply instead in the case of Lagrangian formulations valid
in Minkowski space-time. A solution to this issue has been pointed out in
Ref.\cite{syn}, based on the introduction of a new form of the variational
principle, to be referred to as synchronous action variational principle.
The same type of problem can be posed in the context of continuum
Hamiltonian theories. However, the achievement of such a goal may encounter
potentially far more severe difficulties, unless tensorial canonical
variables are adopted. Such a choice may actually be inhibited \cite%
{pir1,pir2,pir3,dirac1,dirac2,ADM}, for example when asynchronous
variational principles are adopted (see definition given in Ref.\cite{syn}).
Despite these difficulties, the possible solution of the problem appears of
fundamental importance. In fact, it must be noted that the Hamiltonian
formalism has a perspicuous and transparent physical interpretation, even in
the context of covariant formulation, an example of which is provided in Ref.%
\cite{EPJ2}. The symplectic structure of phase-space arises also for
continuum systems, a feature which permits one to construct both local
tensorial Poisson brackets and continuum canonical transformations.

In some respects to be later explained, the viewpoint adopted in this paper
has analogies with the so-called \textit{induced gravity} (or \textit{%
emergent gravity}), namely the conjecture that the geometrical properties of
space-time reveal themselves as a mean field description of microscopic
stochastic or quantum degrees of freedom underlying the classical solution
in the corresponding variational formulation. This is achieved by
introducing a prescribed metric tensor $\widehat{g}_{\mu \nu }$ which is
held constant in the variational principles and therefore also in the
Euler-Lagrange equations. This tensor is to be distinguished from the
variational one $g_{\mu \nu }$. More precisely, $\widehat{g}_{\mu \nu }$
acquires a geometrical interpretation, since by construction it
raises/lowers tensorial indices and prescribes the covariant derivatives. In
this picture, $\widehat{g}_{\mu \nu }$ arises as a macroscopic prescribed
mean field emerging from a background of variational fields $g_{\mu \nu }$,
all belonging to a suitable functional class. This permits to introduce a
new representation for the action functional, depending both on $g_{\mu \nu
} $ and $\widehat{g}_{\mu \nu }$. As shown here, such a feature is found to
be instrumental for the identification of the covariant Hamiltonian
structure associated with the classical gravitational field.

The aim of the paper is to provide a continuum Lagrangian and Hamiltonian
variational formulations for the Einstein equation in the context of a
non-perturbative treatment. Basic feature is the manifest covariance
property of the theory at all levels, whereby all physical quantities
including the canonical variables have a tensorial character. Two remarkable
consequences follow. The first one is that the continuum Hamiltonian
structure is displayed in terms of continuum tensor Poisson brackets.
Second, as a result, the physical interpretation of the canonical variables
and the extremal quantities (in particular in terms of $\widehat{g}_{\mu \nu
}$ and of the extremal Hamiltonian function) clearly emerges.

\bigskip

In detail, the scheme of the paper is as follows. Section II contains the
principles of the axiomatic formulation which defines the theoretical
background for the present study. Section III presents a critical review of
the main features on the Lagrangian formulations of GR given in the
literature. Section IV deals with the definition of a constrained
variational principle (THM.1)\ which allows one to treat the contribution of
the connection fields in the Lagrangian density consistent with the
principles of GR and its logic foundations. In Section V we present a
discussion about some relevant features characterizing the variational
treatments of field theories based on Lagrangian densities. In Section VI
the constrained synchronous variational principle is formulated (THM.2),
which makes possible the use of 4-scalar Lagrangian functions for the
derivation of the Einstein equations instead of Lagrangian densities. In
Section VII the result is extended to the formulation of a generalized
constrained synchronous variational principle (THM.3) with the inclusion of
generalized field velocity expressed by the covariant derivative of the
variational metric tensor. Then, it is shown that this permits to cast the
Einstein equations as Euler-Lagrange equations in manifest-covariant
standard form (Corollary to THM.3). Section VIII presents a
manifest-covariant Hamiltonian formulation of the field equations (THM.4)
carried out in terms of a covariant definition of canonical momenta and a
non-singular Legendre transform. In Section IX the extension of the theory
to the inclusion of the cosmological constant is described (THM.5) and a
characteristic gauge invariance property of the synchronous principle is
pointed out (THM.6). Concluding remarks are then presented in Section X.

\section{Historical background}

A\ common theoretical feature shared by General Relativity (GR) and
classical field theory is the variational character of the fundamental
dynamical laws which identify these disciplines. This concerns both the
representation of the Einstein field equations and of the covariant dynamics
of classical fields as well as of discrete (e.g., point-like or extended
particles) or continuum systems in curved space-time. For definiteness, the
non-vacuum Einstein equations for the metric tensor $g_{\mu \nu }$,\textbf{\
}assumed to have Lorentzian signature\textbf{\ }$(+,-,-,-)$, in the absence
of cosmological constant are given by%
\begin{equation}
R_{\mu \nu }-\frac{1}{2}Rg_{\mu \nu }=\frac{8\pi G}{c^{4}}T_{\mu \nu },
\label{eis}
\end{equation}%
which are assumed to hold in the 4-dimensional space-time\textbf{\ }$D^{4}$,%
\textbf{\ }to be identified with a connected and time-oriented Lorentzian
differentiable manifold $(\mathbf{Q}^{4},g)$. Here $G_{\mu \nu }\equiv
R_{\mu \nu }-\frac{1}{2}Rg_{\mu \nu }$ defines the symmetric Einstein
tensor, while $R_{ik}$ is the Ricci curvature 4-tensor%
\begin{equation}
R_{\mu \nu }=\frac{\partial \Gamma _{\mu \nu }^{\alpha }}{\partial x^{\alpha
}}-\frac{\partial \Gamma _{\mu \alpha }^{\alpha }}{\partial x^{\nu }}+\Gamma
_{\mu \nu }^{\alpha }\Gamma _{\alpha \beta }^{\beta }-\Gamma _{\mu \alpha
}^{\beta }\Gamma _{\nu \beta }^{\alpha },  \label{rirr}
\end{equation}%
with $\Gamma _{\mu \nu }^{\alpha }$ denoting non-tensorial quantities which
identify the so-called (standard) connection functions. Finally, the source
term on the rhs is expressed in terms of the symmetric stress-energy tensor $%
T_{\mu \nu }$. Eq.(\ref{eis}) determines the dynamical equations of $g_{\mu
\nu }\left( r\right) $ to be completed by the metric compatibility condition
of the covariant derivative%
\begin{equation}
\nabla _{\alpha }g^{\mu \nu }=0,  \label{metric-com}
\end{equation}%
which provides a unique relationship between $g_{\mu \nu }$ and the
connection function contained in the covariant derivative operator $\nabla
_{\alpha }$ \cite{LL,wald}. In particular, this yields the definition of the
Christoffel symbols as%
\begin{equation}
\Gamma _{\mu \nu }^{\alpha }=\frac{1}{2}g^{\alpha \beta }\left( \frac{%
\partial g_{\mu \beta }}{\partial x^{\nu }}+\frac{\partial g_{\beta \nu }}{%
\partial x^{\mu }}-\frac{\partial g_{\mu \nu }}{\partial x^{\beta }}\right) .
\label{ks}
\end{equation}

Fundamental issues related to possible alternative construction methods of
Lagrangian variational formulations for the Einstein equation have been
treated extensively in Ref.\cite{syn}. The analysis has permitted to uncover
the existence of a new class of variational approaches for classical
continuum fields, referred to as \textit{synchronous Lagrangian variational
principles. }These have been shown to depart from the customary approach
adopted in the literature and the analogous one developed originally by
Einstein himself for his namesake equation. The key feature of the new
approach, which sets is apart from such literature approaches, lies in the
prescription of the functional setting for the variational classical fields,
whereby the $4-$scalar 4-volume element $d\Omega \equiv d^{4}x\sqrt{-g}$
(with $g$ denoting the determinant of the metric tensor) is held fixed
during arbitrary variations performed on the\ same fields. In contrast,
traditional Lagrangian approaches to be found in the literature are based on
so-called \textit{asynchronous variations}, namely in which $d\Omega $ is
actually suitably varied during the variations of classical fields. Such a
property, as pointed out in Ref.\cite{syn}, has important implications since
it permits to cure deficiencies inherent in the said literature approaches.
These include in particular the property of manifest covariance of the
symbolic Euler-Lagrange equations and the gauge invariance property of the
variational Lagrangians.

Based on these premises, in this paper we want to address the issue of
constructing a covariant Hamiltonian theory of GR. Such a problem must be
set in the proper historical perspective. Indeed, it is well-known that
although in principle from the Lagrangian formulation it is possible to
derive a Hamiltonian formulation, in the case of GR this is not an easy
task. In fact, in the past such a Hamiltonian theory has been approached
based on the principles of constrained dynamics and the related concepts of
primary constraints (in which the constraints depend both on coordinates and
momenta) and secondary constraints (i.e., those which are not primary).
Therefore, in case of Hamiltonian systems they coincide respectively with
the so-called first- and second-class constraint, originally introduced by
Dirac in order to treat singular systems \cite{dirac4,dirac3}. Based on this
approach, some qualitative properties emerge which are problematic. These
are:

1) Lack of manifest covariance of the variational functions which appear
both in the Lagrangian and Hamiltonian formulations. These include for
example the same Lagrangian and Hamiltonian densities as well as the
definition of the canonical momenta. A basic consequence is that the
symbolic Euler-Lagrange equations are not manifestly covariant.

2) Lack of gauge invariance properties for the Hamiltonian theory. This
property is actually inherited from the corresponding Lagrangian
formulation, see Ref.\cite{syn}.

3) Another possible issue concerns the definition of the variational
functional class together with the boundary conditions to be satisfied by
the varied fields.

4) A further deficiency which is in part connected to 3 is the fact that it
is not possible to define boundary conditions for fields which do not have
4-tensor properties. In particular, such boundary conditions only apply in a
given coordinate system. Therefore, they are mutually related by coordinate
transformation. This means that, as a result, if the boundary condition is
such that a 4-tensor is set to zero on the boundary $\partial \mathbb{D}^{4}$%
, then since its modulus is a 4-scalar which is identically zero, the same
modulus condition must apply in any coordinate system.

In order to analyze systematically the previous issue, it is worth starting
from the original approaches based on the Einstein-Hilbert Lagrangian. First
attempts to develop a Hamiltonian theory are those due to Pirani and Schild
(1950 \cite{pir1}), Bergmann et al. (1950 \cite{pir2}) and Pirani et al.
(1952 \cite{pir3}), in which the \textquotedblleft
natural\textquotedblright\ choice was made of using the covariant metric
tensor as the canonical (coordinate) variable. Later, Dirac himself
(1958-1959 \cite{dirac1,dirac2}), adopting the same choice, proposed a
Hamiltonian formulation in terms of a modified Einstein-Hilbert Lagrangian.
All such approaches are based on the adoption of non-tensorial canonical
momenta. In fact, in these methods one actually introduces a separation
between \textquotedblleft space\textquotedblright\ and \textquotedblleft
time\textquotedblright\ components of the field metric and its derivatives.
On the other hand, it must be noted that this operation does not correspond
to the introduction of preferred coordinate systems which can result in the
violation of the intrinsic covariance property of the theory \cite%
{kiri2,kiri}. As a consequence, the final set of variational equations are
actually equivalent to the Einstein equations, so that in this sense the
covariance of the Einstein equations is preserved, although the manifest
covariance for such approaches is not extant.

To key principles of Dirac's Hamiltonian approach is that of singling out
the \textquotedblleft time\textquotedblright\ component of the 4-position
vector in terms of which the generalized velocity is identified with $g_{\mu
\nu ,0}$. This lead him to identify the canonical momentum in terms of the
manifest non-tensorial quantity%
\begin{equation}
\pi _{Dirac}^{\mu \nu }=\frac{\partial L_{EH}}{\partial g_{\mu \nu ,0}},
\label{momentum-dir}
\end{equation}%
where $L_{EH}$ is the Einstein-Hilbert variational Lagrangian density. One
can even argue at this point on the correctness of such an identification of
the canonical momentum. In particular, the definition of the momentum based
on the partial derivative of a tensor can appear ambiguous in the framework
of GR, where only covariant derivatives posses the proper covariance
character.

Another well-known approach to the Hamiltonian formulation of the Einstein
equations is the one developed by Arnowitt, Deser and Misner (1959-1962 \cite%
{ADM}, usually referred to as ADM theory). Detailed presentations of this
issue can be found for example in Refs.\cite{gravi,wald}. There is a
striking analogy with the Dirac approach, in that also in this case the role
of the time coordinate is singled out. The theory provides a 3+1
decomposition of GR which is coordinate invariant but foliation dependent,
in the sense that it relies on a choice of a family of observers according
to which space-time is split into time and a 3-space \cite{alcu}.
Nevertheless, also in this case manifest covariance is lost, specifically
because of the adoption, inherent in the same ADM approach, of non-4$-$%
tensor Lagrangian and Hamiltonian variables. This feature supports the
objection raised by Hawking against the ADM theory, who stated that
\textquotedblleft the split into three spatial dimensions and one time
dimension seems to be contrary to the whole spirit of
Relativity\textquotedblright\ \cite{haw} (see also Ref.\cite{penrose} for
additional critics on the 3+1 decomposition). This point of view seems
physically well-founded. In fact, in the spirit of GR, \textquotedblleft
time\textquotedblright\ and \textquotedblleft space\textquotedblright\
should be treated on equal footing as independent variables. The distinction
among the entries of 4-tensors cannot be longer put on physical basis in GR,
in contrast to what happens in flat space-time. As a result, the special
role attributed to the \textquotedblleft time\textquotedblright\ (or zero)
component with respect to the \textquotedblleft space\textquotedblright\
components does not appear consistently motivated.

The remaining critical points concern issues 2-4 indicated above. In fact
both the Dirac and the ADM approaches rely on the original Einstein-Hilbert
Lagrangian density. As pointed out in Ref.\cite{syn}, this misses the
correct gauge invariance properties required for a variational formulation
in classical field theory. Therefore, also the corresponding Hamiltonian
approaches share the same feature. In addition, it must be remarked that in
both Hamiltonian approaches the variational functional class is not defined
so it is not clear which boundary conditions are to be set on the varied
fields. In this regard we also notice that it seems doubtful even the
possibility of prescribing in a consistent unique way boundary conditions on
non-tensorial fields.

Because of these reasons, it still remains to be ascertained whether
alternative approaches can be envisaged which can ultimately overcome
simultaneously all these problems. The Lagrangian formulation of GR has been
already discussed in Ref.\cite{syn}. This leaves however the problem of the
corresponding Hamiltonian formulations of GR still open.

\section{Physical motivations}

In this section we introduce the physical motivations for the research
program developed in this paper. In this regard a basic preliminary issue to
be first addressed concerns the possible viewpoint denying the very
possibility of a manifest covariant Hamiltonian field theory - in contrast
to the corresponding Lagrangian one - \cite{wald}. The goal of this paper is
actually to prove the falsity of such a statement. However, a first hint at
such a conclusion emerges also from elementary considerations.

The first motivation arises from field theory. In fact in the literature the
problem of the construction of manifest covariant Hamiltonian formulations
for field theories has a long history. In this reference, the earliest
contributions date back to the pioneering work by De Donder (1930 \cite%
{donder})\ and Weyl (1935 \cite{weyl}). More recently, the subject has been
treated in Refs.\cite{sym1,sym2,sym3,sym4,sym5,sym6,sym7,sym8,sym9}, where
such an approach is usually referred to as \textquotedblleft
multi-symplectic\textquotedblright\ or \textquotedblleft poly-symplectic
field theory\textquotedblright . The general method underlying this approach
can be best illustrated by considering the example-case of a scalar field $%
\phi \equiv \phi \left( r^{\mu }\right) $ in flat space-time, being the
treatment of tensorial fields analogous \cite{sym9}. For definiteness let us
consider a Minkowski space-time $\left( \mathbf{M}^{4},\eta \right) $ with
signature $(+,-,-,-)$. The corresponding variational Lagrangian density $%
L_{\phi }$ becomes a 4-scalar, since $\sqrt{-g}=1$ identically. The
Lagrangian is assumed to exhibit a functional dependence of the form $%
L_{\phi }=L_{\phi }\left( \phi ,\partial _{\mu }\phi ,r^{\mu }\right) $,
namely to depend on the field $\phi $, its first partial derivatives and
possibly also explicitly on the position 4-vector $r^{\mu }$. As a result,
the action integral is written as%
\begin{equation}
S_{\phi }=\int_{M^{4}}d^{4}rL_{\phi }\left( \phi ,\partial _{\mu }\phi
,r^{\mu }\right) .
\end{equation}%
Variation of $S_{\phi }$ in a suitable functional class in which the value
of $\phi $ is prescribed on a suitable fixed boundary of $M^{4}$ then
provides the Euler-Lagrange equation%
\begin{equation}
\frac{\partial }{\partial x^{\mu }}\frac{\partial L_{\phi }}{\partial \left(
\partial _{\mu }\phi \right) }-\frac{\partial L_{\phi }}{\partial \phi }=0,
\label{el-fi}
\end{equation}%
which determines the behavior of $\phi \left( r^{\mu }\right) $.

The corresponding Hamiltonian formulation is obtained by identifying the
conjugate momentum $\pi _{\phi }^{\mu }$ in terms of the 4-vector%
\begin{equation}
\pi _{\phi }^{\mu }=\frac{\partial L_{\phi }}{\partial \left( \partial _{\mu
}\phi \right) },
\end{equation}%
so that, provided\ $L_{\phi }$ is regular, the 4-scalar Hamiltonian density $%
\mathcal{H}_{\phi }=\mathcal{H}_{\phi }\left( \phi ,\pi _{\phi }^{\mu
},r^{\mu }\right) $ is introduced by means of the Legendre transform%
\begin{equation}
\mathcal{H}_{\phi }=\pi _{\phi }^{\mu }\partial _{\mu }\phi -L_{\phi },
\end{equation}%
which is usually referred to as the \textquotedblleft De
Donder-Weyl\textquotedblright\ Hamiltonian density for the scalar field $%
\phi $. The set of first-order Hamilton equations corresponding to the
Euler-Lagrange equation (\ref{el-fi}) which define the continuum Hamiltonian
system are provided by the covariant PDEs%
\begin{eqnarray}
\frac{\partial \mathcal{H}_{\phi }}{\partial \pi _{\phi }^{\mu }} &=&\frac{%
\partial \phi }{\partial r^{\mu }},  \label{DW-h1} \\
\frac{\partial \mathcal{H}_{\phi }}{\partial \phi } &=&-\frac{\partial \pi
_{\phi }^{\mu }}{\partial r^{\mu }}.  \label{DW-h2}
\end{eqnarray}%
Remarkably, in such a framework the variational principle meets all the
requirements indicated above. In particular, one notices that the canonical
variables $\left( r^{\mu },\pi _{\phi }^{\mu }\right) $ are both 4-vectors
and the Hamiltonian density is a 4-scalar, so that the Hamilton equations (%
\ref{DW-h1}) and (\ref{DW-h2}) are manifestly covariant. Finally, the theory
is manifestly gauge invariant.

The second motivation is provided by the analogy with the variational
description of relativistic particle dynamics, which can be cast both in
Lagrangian and Hamiltonian forms by preserving manifest covariance. The
construction of the Hamiltonian formulation is a direct consequence of the
synchronous Lagrangian variational principle (see Ref.\cite{syn}). In fact,
let us consider a point-like particle having rest mass $m_{o}$, charge $%
q_{o} $ and proper time $s$, so that the corresponding 4-velocity is $u^{\mu
}\left( s\right) =\frac{dr^{\mu }\left( s\right) }{ds}$, while the line
element is given by%
\begin{equation}
ds^{2}=g_{\mu \nu }(r)dr^{\nu }(s)dr^{\mu }(s).  \label{ds2}
\end{equation}%
Here the metric tensor $g_{\mu \nu }\left( r\right) $ and the Faraday tensor
$F_{\mu \nu }=\partial _{\mu }A_{\nu }-\partial _{\nu }A_{\mu }$ of the
external EM fields are considered prescribed, with $A_{\mu }$ denoting the
4-vector potential. The Lagrangian functional for this system is given by%
\begin{equation}
S_{pS}(r^{\mu },u^{\mu
})=-\int\nolimits_{s_{1}}^{s_{2}}dsL_{pS}\left( r^{\mu }\left(
s\right) ,\overset{\cdot }{r}^{\mu }\left( s\right) ,u^{\mu
}\left( s\right) \right) ,
\end{equation}%
where $\overset{\cdot }{r}^{\mu }\equiv \frac{dr^{\mu }}{ds}$ and $L_{pS}$
is the 4-scalar Lagrangian%
\begin{equation}
L_{pS}\left( r^{\mu }\left( s\right) ,\overset{\cdot }{r}^{\mu }\left(
s\right) ,u^{\mu }\left( s\right) \right) \equiv \left( u_{\mu }\left(
s\right) +qA_{\mu }\left( r\left( s\right) \right) \right) \frac{dr^{\mu
}\left( s\right) }{ds}-\frac{1}{2}u^{\mu }\left( s\right) u_{\mu }\left(
s\right) .
\end{equation}%
Here $q\equiv \frac{q_{o}}{m_{o}c^{2}}$ is the normalized charge. The virtue
of this variational formulation is the adoption of superabundant variables,
so that the kinematic constraint (mass-shell condition) $u^{\mu }\left(
s\right) u_{\mu }\left( s\right) =1$ acts only on the extremal function,
while the line element $ds$ is by construction expressed by Eq.(\ref{ds2})
in terms of the extremal curve $r^{\mu }\left( s\right) $.

Then, after introducing the customary Legendre transformation, one recovers
the Hamiltonian variational formulation which is provided in terms of the
functional%
\begin{equation}
S_{pS}^{H}(r^{\mu },p^{\mu
})=\int\nolimits_{s_{1}}^{s_{2}}ds\left[ p_{\mu }\left( s\right)
\frac{dr^{\mu }\left( s\right) }{ds}-H_{pS}\right] ,
\end{equation}%
where%
\begin{equation}
H_{pS}=\frac{1}{2}\left( p_{\mu }-qA_{\mu }\right) \left( p^{\mu }-qA^{\mu
}\right)  \label{super-man}
\end{equation}%
is the 4-scalar Hamiltonian function and $p_{\mu }\equiv u_{\mu }+qA_{\mu }$
is the canonical 4-momentum. The corresponding variational principle is
realized in terms of the synchronous Hamilton variational principle.
Following Ref.\cite{syn}, we introduce the synchronous functional class%
\begin{equation}
\left\{ r^{\mu },p^{\mu }\right\} _{S}=\left\{
\begin{array}{c}
r^{\mu }(s),p^{\mu }\left( s\right) \in C^{2}\left(
\mathbb{R}
\right) \\
\delta \left( ds\right) =0 \\
r^{\mu }(s_{k})=r_{k}^{\mu },\ k=1,2 \\
p^{\mu }(s_{k})=p_{k}^{\mu },\ k=1,2%
\end{array}%
\right\} ,
\end{equation}%
where $\delta $ denotes the synchronous variation operator. This is defined
with respect to the extremal quantities, namely%
\begin{eqnarray}
\delta r^{\mu }(s) &=&r^{\mu }(s)-r_{extr}^{\mu }(s),  \label{44} \\
\delta p^{\mu }(s) &=&p^{\mu }(s)-p_{extr}^{\mu }(s),  \label{44-bis}
\end{eqnarray}%
where $r_{extr}^{\mu }(s)$ and $p_{extr}^{\mu }(s)$ identify the extremal
phase-space curves. Here, $r^{\mu }(s)$ and $p^{\mu }(s)$ are considered
independent, so that $\delta r^{\mu }(s)$ and $\delta p^{\mu }(s)$ are
independent too. In this setting, the variational principle is realized in
terms of the synchronous variation of the functional $S_{pS}^{H}(r^{\mu
},p^{\mu })$, namely the corresponding Frechet derivative%
\begin{equation}
\delta S_{pS}^{H}(r^{\mu },p^{\mu })\equiv \left. \frac{d}{d\alpha }\Psi
(\alpha )\right\vert _{\alpha =0}=0,  \label{delta-SS}
\end{equation}%
to hold for arbitrary independent displacements $\delta r^{\mu }(s)$ and $%
\delta p^{\mu }(s)$. Notice that here $\Psi (\alpha )$ is the smooth real
function $\Psi (\alpha )=S_{pS}(r^{\mu }+\alpha \delta r^{\mu },p^{\mu
}+\alpha \delta p^{\mu })$, being $\alpha \in \left] -1,1\right[ $ to be
considered independent of $r^{\mu }(s)$, $p^{\mu }\left( s\right) $ and $s$,
so that the corresponding variational derivatives are%
\begin{eqnarray}
\frac{\delta S_{pS}^{H}(r^{\mu },p^{\mu })}{\delta r^{\mu }(s)} &\equiv &-%
\frac{D}{Ds}p_{\mu }-\frac{\partial H_{pS}}{\partial r^{\mu }}=0,
\label{ss1} \\
\frac{\delta S_{pS}^{H}(r^{\mu },p^{\mu })}{\delta p_{\mu }(s)} &\equiv &%
\frac{dr^{\mu }\left( s\right) }{ds}-\frac{\partial H_{pS}}{\partial p_{\mu }%
}=0,  \label{ss2}
\end{eqnarray}%
which identify the corresponding 1-body Hamiltonian system. In Eq.(\ref{ss1}%
) $\frac{D}{Ds}$ denotes the covariant derivative acting on $p_{\mu }$ \cite%
{LL}. In the literature (see for example Ref.\cite{gravi}) Eq.(\ref%
{super-man}) is usually referred to as \textquotedblleft
super-Hamiltonian\textquotedblright , with Eqs.(\ref{ss1}) and (\ref{ss2})
as super-Hamiltonian equations. Nevertheless, we notice that the same
equations exhibit the standard Hamiltonian structure \cite{EPJ2}. This means
that the set $\left\{ H_{pS},\mathbf{y}\equiv \left( r^{\mu },p^{\mu
}\right) \right\} $ defines in a proper sense a Hamiltonian system, even if
it is represented in terms of a superabundant canonical state. Its
remarkable property in this respect is that, once the constraint (\ref{ds2})
on the line element to hold only for the extremal curves has been set, no
further constraints are required on the extremal canonical state $\overline{%
\mathbf{y}}$. Indeed, such constraints are now identically fulfilled simply
as a consequence of the Hamilton equations (\ref{ss1}) and (\ref{ss2}).
These conclusions depart from the well-known Dirac generating-function
formalism (DGF) where the canonical variables do not have a tensorial
character, although the property of covariance (albeit not of manifest
covariance) of the related Hamilton equations is still fulfilled (see
related discussion in Ref.\cite{EPJ3}).

Furthermore, it is interesting to notice the basic properties of the
synchronous Hamiltonian variational principle (\ref{delta-SS}), which
involve: 1) coordinates and momenta which are 4-vectors;\ 2)\ the
Hamiltonian function $H_{pS}$ and the action functional $S_{pS}^{H}(r^{\mu
},p^{\mu })$ which are a 4-scalars; 3)\ the manifestly-covariant Hamilton
equations. In addition, it must be noticed that the Hamiltonian $H_{pS}$ has
a special feature, namely its extremal value is a constant equal to $1/2$.
Such a theory is of general validity and relies on the adoption of Hamilton
variational principle, which holds even for the treatment of the non-local
interaction occurring in the EM\ radiation-reaction problem. A theory of
this type has been recently established in Refs.\cite%
{EPJ1,EPJ2,EPJ3,EPJ4,EPJ5,EPJ6,EPJ7,EPJ8}.

These considerations suggest the obvious conjecture that the same program
can actually be worked out in curved space-time, and in particular for the
gravitational field in such a way to determine a covariant Hamiltonian
variational formulation for the Einstein equations.

\section{Modified synchronous Lagrangian variational principle}

In view of the considerations outlined above, let us now pose the problem of
constructing a modified version of the synchronous Lagrangian variational
principle for the Einstein equation reported in Ref.\cite{syn}. The issue,
as will be clarified below, is actually propedeutic for the problem
considered in this paper about the Hamiltonian formulation of GR. Starting
point is the synchronous Lagrangian variational principle presented in Ref.%
\cite{syn}, recalled here for better clarity. We adopt the same notation of
Ref.\cite{syn}.

To ease the presentation we consider the case of vacuum Einstein equations.
The extension to the case of non-vacuum equations in the presence of EM and
external matter sources is given in Ref.\cite{syn}. As also shown in THM.2
of the same reference, a synchronous variational principle of the type%
\begin{equation}
\delta S_{1}\left( Z,\widehat{Z}\right) =0
\end{equation}%
holds, which applies for arbitrary variations of the variational fields $Z$
and with the operator $\delta $ denoting the \emph{synchronous variation
operator} (see Ref.\cite{syn}). By construction, this acts in such a way
that both%
\begin{eqnarray}
\delta \widehat{Z} &\equiv &0,  \label{constraint-eq} \\
\delta Z_{extr} &\equiv &0,
\end{eqnarray}%
with $\widehat{Z}$ and\ $Z_{extr}$ identifying respectively the prescribed
and the extremal tensor fields. In particular, $Z_{extr}$ is defined as the
solution of the boundary-value problem associated with the Euler-Lagrange
equations. Instead, when acting on any other arbitrary variational function $%
Z$ different from both $\widehat{Z}$ and $Z_{extr}$, the synchronous
variation is defined as%
\begin{equation}
\delta Z\left( r\right) =Z\left( r\right) -Z_{extr}\left( r\right) .
\label{new-orig}
\end{equation}%
This requires identifying $\Psi (\alpha )=S_{1}(Z+\alpha \delta Z,\widehat{Z}%
)$, where the action functional is%
\begin{equation}
S_{1}\left( Z,\widehat{Z}\right) =\int_{\widehat{D}^{4}}d\Omega L_{1}\left(
Z,\widehat{Z}\right) .  \label{s1-diz}
\end{equation}%
In this treatment, $Z$, $\delta Z$ and $\widehat{Z}$ denote respectively the
variational, the variation and the prescribed fields, the latter being held
fixed during synchronous variations, so that identically $\delta \widehat{Z}%
\left( r\right) =0$ (see discussion in Ref.\cite{syn}). Thus, here $L_{1}$
is the 4-scalar variational Lagrangian density%
\begin{equation}
L_{1}\left( Z,\widehat{Z}\right) \equiv -\frac{c^{3}}{16\pi G}g^{\mu \nu }%
\widehat{R}_{\mu \nu }h\left( Z,\widehat{Z}\right) ,  \label{lxc}
\end{equation}%
where $h\left( Z,\widehat{Z}\right) $ is the 4-scalar\ multiplicative factor%
\begin{equation}
h\left( Z,\widehat{Z}\right) \equiv \left( 2-\frac{1}{4}g^{\alpha \beta
}g_{\alpha \beta }\right) ,  \label{hh}
\end{equation}%
and $\widehat{R}_{\mu \nu }$ is defined according to Eq.(\ref{rirr}) and is
evaluated for $g_{\mu \nu }\left( r\right) =\widehat{g}_{\mu \nu }\left(
r\right) $, while here $g^{\alpha \beta }g_{\alpha \beta }\neq \delta
_{\alpha }^{\alpha }$ for variational curves (see below). Here, both $Z$ and
$\widehat{Z}$ belong to the functional class%
\begin{equation}
\left\{ Z\right\} _{E-S}\equiv \left\{
\begin{array}{c}
Z_{1}\left( r\right) \equiv g_{\mu \nu }\left( r\right) \\
\widehat{Z}_{1}\left( r\right) \equiv \widehat{g}_{\mu \nu }\left( r\right)
\\
\widehat{Z}_{2}\left( r\right) \equiv \widehat{R}_{\mu \nu }\left( r\right)
\\
Z\left( r\right) ,\widehat{Z}\left( r\right) \in C^{k}\left( \mathbb{D}%
^{4}\right) \\
g_{\mu \nu }\left( r\right) |_{\partial \mathbb{D}^{4}}=g_{\mu \nu \mathbb{D}%
}\left( r\right) \\
\delta \widehat{Z}\left( r\right) =\delta Z_{extr}\left( r\right) =0 \\
\delta \left( d\Omega \right) =0 \\
g_{\mu \nu }=g^{\alpha \beta }\widehat{g}_{\alpha \mu }\widehat{g}_{\beta
\nu } \\
g^{\alpha \beta }=\widehat{g}^{\alpha \mu }\widehat{g}^{\beta \nu }g_{\mu
\nu } \\
\nabla _{\alpha }\equiv \widehat{\nabla }_{\alpha }%
\end{array}%
\right\} ,  \label{Z-E-S}
\end{equation}%
where $k\geq 3$ and all the fields $\left\{ g_{\mu \nu }\left( r\right) ,%
\widehat{g}_{\mu \nu }\left( r\right) ,\widehat{R}_{\mu \nu }\left( r\right)
\right\} $\textbf{\ }are by assumption symmetric in the indices $\mu ,\nu $.
We stress that in $\left\{ Z\right\} _{E-S}$ it is assumed that the
variational metric tensor $Z_{1}\left( r\right) \equiv g_{\mu \nu }\left(
r\right) $ does not raise/lower indices. Instead, the covariant varied
function $g_{\alpha \beta }\left( r\right) $ must be transformed into the
corresponding contravariant representation $g^{\alpha \beta }\left( r\right)
$\ by means of the fixed metric tensor $\widehat{g}_{\alpha \beta }$ only,
namely $g_{\alpha \beta }=\widehat{g}_{\alpha \mu }\widehat{g}_{\beta \nu
}g^{\mu \nu }$. Furthermore, $\widehat{\nabla }_{\alpha }$ denotes the
covariant derivative expressed in terms of $\widehat{g}_{\alpha \beta }$.

Based on these premises, the following modified form of the Lagrangian
action principle can be established.

\bigskip

\textbf{THM.1 - Extended form of the synchronous Lagrangian variational
principle}

\emph{Given validity of THM.2\ in Ref.\cite{syn}, let us introduce in the
action functional the 4-scalar contribution to the Lagrangian variational
density}%
\begin{equation}
\Delta L_{1}\left( Z,\widehat{\nabla }_{\mu }Z,\widehat{Z}\right) \equiv
\frac{1}{2}\kappa \widehat{\nabla }^{k}g_{\mu \nu }\widehat{\nabla }%
_{k}g^{\mu \nu }h\left( Z,\widehat{Z}\right) ,  \label{delta-L1}
\end{equation}%
\emph{where }$\widehat{\nabla }_{k}$\emph{\ denotes the covariant derivative
operator expressed by the Christoffel symbols evaluated in terms of the
prescribed metric tensor }$\widehat{g}_{\mu \nu }$\emph{\ which is held
constant during synchronous variations. Moreover the same tensor is used to
raise/lower indices in }$\Delta L_{1}\left( Z,\widehat{\nabla }_{\mu }Z,%
\widehat{Z}\right) $\emph{, so that in particular }$\widehat{\nabla }%
^{k}\equiv \widehat{g}^{ik}\widehat{\nabla }_{i}$\emph{. Finally, }$\kappa $%
\emph{\ is a suitable 4-scalar constant. Then, the following propositions
hold:}

T1$_{1})$\emph{\ Denoting by }$\Delta S_{1}\left( Z,\widehat{Z}\right) $
\emph{the corresponding action functional, one finds that the synchronous
variation of }$\Delta S_{1}\left( Z,\widehat{Z}\right) $ \emph{performed in
the class }$\left\{ Z\right\} _{E-S}$\emph{\ and evaluated for }$g_{\mu \nu
}=\widehat{g}_{\mu \nu }$ \emph{is identically satisfied for arbitrary
variations }$\delta g^{\mu \nu }$\emph{\ in the same functional class.}

T1$_{2})$\emph{\ The dimensional constant }$\kappa $\emph{\ can always be
expressed\ as}%
\begin{equation}
\kappa =\frac{c^{3}}{16\pi G},
\end{equation}%
\emph{namely the dimensional factor which appears in the Einstein-Hilbert
functional.}

T1$_{3})$ \emph{The variational Lagrangian density in the action functional
can always be identified with}%
\begin{eqnarray}
L_{2}\left( Z,\widehat{\nabla }_{\mu }Z,\widehat{Z}\right) &\equiv
&L_{1}\left( Z,\widehat{Z}\right) +\Delta L_{1}\left( Z,\widehat{\nabla }%
_{\mu }Z,\widehat{Z}\right)  \notag \\
&=&-\kappa \left[ g^{\mu \nu }\widehat{R}_{\mu \nu }-\frac{1}{2}\widehat{%
\nabla }^{k}g_{\mu \nu }\widehat{\nabla }_{k}g^{\mu \nu }\right] h\left( Z,%
\widehat{Z}\right) ,  \label{L4}
\end{eqnarray}%
\emph{so that the Lagrangian functional becomes finally}%
\begin{equation}
S_{2}\left( Z,\widehat{Z}\right) =\int_{\widehat{D}^{4}}d\Omega L_{2}\left(
Z,\widehat{\nabla }_{\mu }Z,\widehat{Z}\right) .
\label{esse-4-gen-functional}
\end{equation}

T1$_{4})$ \emph{The symbolic Euler-Lagrangian equations corresponding to the
variational Lagrangian density }$L_{2}\left( Z,\widehat{\nabla }_{\mu }Z,%
\widehat{Z}\right) $ \emph{take the form}%
\begin{equation}
\widehat{\nabla }_{i}\frac{\partial L_{2}\left( Z,\widehat{\nabla }_{\mu }Z,%
\widehat{Z}\right) }{\partial \left( \widehat{\nabla }_{i}g^{\mu \nu
}\right) }-\frac{\partial L_{2}\left( Z,\widehat{\nabla }_{\mu }Z,\widehat{Z}%
\right) }{\partial g^{\mu \nu }}=0,  \label{EL-thm1}
\end{equation}%
\emph{and are therefore manifestly covariant. Explicit evaluation gives}%
\begin{equation}
\widehat{\nabla }_{i}\left[ -h\widehat{\nabla }^{i}g_{\mu \nu }\right] -%
\frac{\partial }{\partial g^{\mu \nu }}\left[ g^{\mu \nu }\widehat{R}_{\mu
\nu }h\right] +\frac{1}{2}\widehat{\nabla }^{k}g_{\mu \nu }\widehat{\nabla }%
_{k}g^{\mu \nu }\frac{\partial }{\partial g^{\mu \nu }}h=0.  \label{prova}
\end{equation}

T1$_{5})$ \emph{If the solution }$\overline{g}^{\mu \nu }\left( r\right) $
\emph{of Eq.(\ref{EL-thm1}), which shall be denoted as extremal curve of the
Lagrangian action }$S_{2}\left( Z,\widehat{Z}\right) $\emph{, is identified
with the prescribed metric tensor }$\widehat{g}^{\mu \nu }\left( r\right) $%
\emph{, then Eq.(\ref{prova}) coincides with the vacuum Einstein equation.}

\emph{Proof -- }To prove T1$_{1}$, let us evaluate the synchronous
variational derivative%
\begin{equation}
\frac{\delta \Delta S_{1}\left( Z,\widehat{Z}\right) }{\delta g^{\mu \nu }}=%
\frac{1}{4}g_{\mu \nu }\kappa \widehat{\nabla }^{k}g_{\alpha \beta }\widehat{%
\nabla }_{k}g^{\alpha \beta }-\kappa \widehat{\nabla }_{k}\left[ h\widehat{%
\nabla }^{k}g_{\alpha \beta }\right] =0.
\end{equation}%
If $g_{\alpha \beta }$ is identified with the prescribed metric tensor $%
\widehat{g}_{\alpha \beta }$, by construction the lhs of such an equation
vanishes identically. To prove the statement T1$_{2}$ it is sufficient to
notice that, from a dimensional analysis it follows that $\left[ \kappa %
\right] =\left[ \frac{c^{3}}{G}\right] $, so that one can choose the
customary coefficient of the Einstein Lagrangian density. This permits one
to represent the total Lagrangian in terms of the same coefficient $\kappa $%
. In addition, the normalization coefficient $\frac{1}{2}$\ is chosen in
analogy with the usual coefficient carried by kinetic terms in
field/particle Lagrangians and for convenience with the subsequent
Hamiltonian formulation (see for example Ref.\cite{wald}). Finally,
proposition T1$_{2}$ is an immediate consequence of T1$_{1}$ and T1$_{2}$.
To prove T1$_{4}$ one simply needs to evaluate the variational derivative of
the functional $S_{2}\left( Z,\widehat{Z}\right) $. It follows that%
\begin{equation}
\frac{\delta S_{2}\left( Z,\widehat{Z}\right) }{\delta g^{\mu \nu }}=-%
\widehat{\nabla }_{i}\frac{\partial L_{2}\left( Z,\widehat{\nabla }_{\mu }Z,%
\widehat{Z}\right) }{\partial \left( \widehat{\nabla }_{i}g^{\mu \nu
}\right) }+\frac{\partial L_{2}\left( Z,\widehat{\nabla }_{\mu }Z,\widehat{Z}%
\right) }{\partial g^{\mu \nu }}=0,
\end{equation}%
which completes the proof. Finally, the proof of proposition T1$_{5}$
follows immediately by identifying $g^{\mu \nu }\left( r\right) $ with $%
\widehat{g}^{\mu \nu }\left( r\right) $ and recalling that by construction $%
h\left( \widehat{g}^{\mu \nu }\left( r\right) \right) =1$, while $\widehat{%
\nabla }_{\alpha }\widehat{g}^{\mu \nu }\left( r\right) =0$.

\textbf{Q.E.D.}

\bigskip

The following remarks must be added:

1)\ Based on the results of Ref.\cite{syn}, one can show that THM.1 above
can be readily extended to treat the case of non-vacuum Einstein equations.
In particular, this includes the Maxwell equations and the treatment of
classical source matter.

2) We stress that the adoption in the variational Lagrangian of the
covariant derivative $\widehat{\nabla }_{i}$ expressed in terms of the same $%
\widehat{g}_{\alpha \beta }$ is actually consistent with the adoption of the
constrained synchronous variational principle. This warrants for example
that, since $d\Omega $ is held fixed in terms of $\widehat{g}_{\alpha \beta
} $, the integration by part in terms of $\widehat{\nabla }_{i}$ works in
the customary way as is permitted by the Gauss theorem.

3) The variational principle in terms of $S_{2}\left( Z,\widehat{Z}\right) $
exhibits the property of manifest covariance.

4)\ The property of gauge invariance, in the sense pointed out in Ref.\cite%
{syn}, is also warranted.

5)\ The contribution $\Delta L_{1}\left( Z,\widehat{\nabla }_{\mu }Z,%
\widehat{Z}\right) $ considered above can be viewed as representing a new
type of gauge invariance, since its extremal value vanishes together with
its functional derivatives when evaluated for the extremal fields. As a
consequence its introduction does not affect in any way, by construction,
the validity of GR and the Einstein equations in particular.

6) Finally, it follows that the Lagrangian differential form and the
corresponding action functional can be defined up to an arbitrary gauge
contribution\ of the form%
\begin{equation}
\Delta S(\widehat{Z})=\int_{\widehat{D}^{4}}d\Omega K(\widehat{Z},r),
\label{delta-esse}
\end{equation}%
where $K$ denotes an arbitrary 4-scalar which depends on prescribed fields
and possibly also on the position 4-vector. This implies that the Lagrangian
$L_{2}\left( Z,\widehat{Z}\right) $ can be equivalently represented as%
\begin{equation}
L_{2-G}\left( Z,\widehat{\nabla }_{\mu }Z,\widehat{Z}\right) =L_{2}\left( Z,%
\overline{\nabla }_{\mu }Z,\overline{Z}\right) +G(\widehat{Z},r),
\end{equation}%
where for example the gauge $G(\widehat{Z},r)$ can be set so that identically%
\begin{equation}
L_{2-G}\left( Z,\widehat{\nabla }_{\mu }Z,\widehat{Z}\right) =0.
\label{L-gauge0}
\end{equation}

As a final comment, we stress that, despite the validity of proposition T1$%
_{5}$, $\widehat{g}_{\alpha \beta }$ and the extremal value of $\overline{g}%
_{\alpha \beta }$, namely the solution of the corresponding Euler-Lagrange
equations given above, remain in principle independent. For the same reason,
the boundary-value functions $g_{\mu \nu \mathbb{D}}\left( r\right) $ should
be regarded as independent from the corresponding ones holding for $\widehat{%
g}_{\alpha \beta }$. The previous considerations notwithstanding, the
identification $\widehat{g}_{\alpha \beta }=\overline{g}_{\alpha \beta }$
remains mandatory for recovering the classical form of the Einstein equation.

\bigskip

The previous conclusions lead us to the following basic result.

\bigskip

\textbf{Corollary 1 to THM.1 - Necessary condition for the solution of the
Einstein equation}

\emph{Given validity of THM.1, it follows that }$\widehat{g}_{\mu \nu
}\left( r\right) $ \emph{is a particular solution of the vacuum Einstein
equation subject to the boundary conditions}%
\begin{equation}
\widehat{g}_{\mu \nu }\left( r\right) |_{\partial \mathbb{D}^{4}}=g_{\mu \nu
\mathbb{D}}\left( r\right)  \label{corollary-bc}
\end{equation}%
\emph{iff it satisfies the boundary-value problem represented by Eq.(\ref%
{prova}) and the same boundary conditions. Then, necessarily the extremal
curve }$g_{\mu \nu }\left( r\right) $ \emph{solution of the same equation in
the functional class }$\left\{ Z\right\} _{E-S}$\emph{\ coincides
identically with }$\widehat{g}_{\mu \nu }\left( r\right) $.

\emph{Proof -- }If $\overline{g}_{\mu \nu }\left( r\right) $ is a particular
solution of the vacuum Einstein equation subject to the boundary conditions (%
\ref{corollary-bc}), then it follows by construction that, since the
covariant derivative $\widehat{\nabla }_{k}\widehat{g}^{\mu \nu }=0$
identically, $\widehat{g}_{\mu \nu }\left( r\right) $ satisfies also Eq.(\ref%
{prova}) and this coincides with the Einstein equation. Viceversa, if $%
\widehat{g}_{\mu \nu }\left( r\right) $ satisfies the boundary-value problem
represented by Eqs.(\ref{prova}) and (\ref{corollary-bc}), then again due to
the vanishing of $\widehat{\nabla }_{k}\widehat{g}^{\mu \nu }$, it obeys
also to the Einstein equation. Finally, due to the uniqueness of the
solutions of the boundary-value problem associated with the Einstein
equation, if $\widehat{g}_{\mu \nu }\left( r\right) $\ satisfies the same
problem, necessarily the extremal curve $\overline{g}_{\mu \nu }\left(
r\right) $ coincides with $\widehat{g}_{\mu \nu }\left( r\right) $.

\textbf{Q.E.D.}

\section{Cosmological constant and non-vacuum Lagrangians}

In this section we analyze how THM.1 can be extended to the treatment of the
following two cases:

1)\ The presence of a non-vanishing cosmological constant $\Lambda $.

2)\ The presence of non-vanishing source matter fields, which determine the
non-vacuum Einstein equation in terms of a stress-energy tensor $T_{\mu \nu
} $.

The treatment is based on the scheme developed in Ref.\cite{syn} for
synchronous Lagrangian variational principles. The solution to points 1 and
2 is provided by the following propositions. The straightforward proofs are
left to the reader.

\bigskip

\textbf{Corollary 2 to THM.1 - Cosmological constant}

\emph{Given validity of THM.1, in the presence of a non-vanishing
cosmological constant }$\Lambda $ \emph{the variational Lagrangian density
becomes}%
\begin{equation}
L_{2\Lambda }\left( Z,\widehat{\nabla }_{\mu }Z,\widehat{Z}\right)
=L_{2}\left( Z,\widehat{\nabla }_{\mu }Z,\widehat{Z}\right) +L_{\Lambda
}\left( Z,\widehat{Z}\right) ,
\end{equation}%
\emph{where }$L_{2}\left( Z,\widehat{\nabla }_{\mu }Z,\widehat{Z}\right) $
\emph{is given by Eq.(\ref{L4}), while }$L_{\Lambda }\left( Z,\widehat{Z}%
\right) $ \emph{is defined as}%
\begin{equation}
L_{\Lambda }\left( Z,\widehat{Z}\right) =-2k\Lambda h\left( Z,\widehat{Z}%
\right) .  \label{L-lambda}
\end{equation}%
\emph{Then THM.1 holds also for the modified Lagrangian }$L_{2\Lambda
}\left( Z,\overline{Z}\right) $\emph{\ in the functional class }$\left\{
Z\right\} _{E-S}$\emph{. The term }$L_{\Lambda }\left( Z,\widehat{Z}\right) $%
\emph{\ gives the contribution}%
\begin{equation}
\frac{\partial L_{\Lambda }\left( Z,\widehat{Z}\right) }{\partial g^{\mu \nu
}}=k\Lambda g_{\mu \nu }.
\end{equation}%
\emph{As a consequence, the Euler-Lagrange equation evaluated for }$%
\overline{g}_{\mu \nu }=g_{\mu \nu }$\emph{\ is}%
\begin{equation}
\overline{R}_{\mu \nu }-\frac{1}{2}\left( \overline{g}^{ik}\overline{R}%
_{ik}\right) \overline{g}_{\mu \nu }+\Lambda \overline{g}_{\mu \nu }=0.
\end{equation}

\bigskip

\textbf{Corollary 3 to THM.1 - Non-vacuum Einstein equations}

\emph{Given validity of THM.1, in the presence of source matter fields, the
variational Lagrangian density becomes}%
\begin{equation}
L_{2F}\left( Z,\widehat{\nabla }_{\mu }Z,\widehat{Z}\right) =L_{2}\left( Z,%
\widehat{\nabla }_{\mu }Z,\widehat{Z}\right) +L_{1F}\left( Z,\widehat{Z}%
\right) ,
\end{equation}%
\emph{where }$L_{2}\left( Z,\widehat{\nabla }_{\mu }Z,\widehat{Z}\right) $
\emph{is given by Eq.(\ref{L4}), while }$L_{1F}\left( Z,\widehat{Z}\right) $
\emph{is defined according to THM.3 in Ref.\cite{syn} as }%
\begin{equation}
L_{1F}\left( Z,\widehat{Z}\right) =L_{F}\left( Z,\widehat{Z}\right) h\left(
Z,\widehat{Z}\right) ,  \label{L1-effe}
\end{equation}%
\emph{with }$L_{F}\left( Z,\widehat{Z}\right) $ \emph{to be suitably
identified with the appropriate EM and matter sources (see Ref.\cite{syn}).
As a consequence, the stress-energy tensor, to be evaluated for }$\widehat{g}%
_{\mu \nu }=\overline{g}_{\mu \nu }$\emph{, which enters the extremal
non-vacuum Einstein equation is defined as}%
\begin{equation}
T_{\mu \nu }\left( r\right) =-2\frac{\partial L_{1F}\left( Z,\overline{Z}%
\right) }{\partial g^{\mu \nu }}+g_{\mu \nu }L_{1F}\left( Z,\overline{Z}%
\right) .
\end{equation}%
\emph{The corresponding synchronous Lagrangian variational principle defined
in THM.1 and evaluated in terms of }$L_{2F}\left( Z,\widehat{\nabla }_{\mu
}Z,\widehat{Z}\right) $\emph{\ yields as extremal equation, when letting }$%
\overline{g}_{\mu \nu }=g_{\mu \nu }$\emph{, the non-vacuum Einstein
equation (\ref{eis}).}

\bigskip

The validity of the two corollaries permits to retain the full form of the
non-vacuum Einstein equation, including also the cosmological constant in
the subsequent developments. Here we stress that in view of Eq.(\ref%
{L-gauge0}), also in the case of non-vacuum theories, the total Lagrangian
density can be prescribed in such a way that its extremal value (obtained
letting $Z=\overline{Z}$) vanishes identically.

\section{Axiomatic formulation to continuum Hamiltonian systems}

In view of the considerations outlined above, let us now pose the problem of
formulating an axiomatic approach in which all the physical requirements
pointed out are included by means of suitable axioms. In particular, this
concerns the extension of the theory outlined by De Donder and Weyl to the
setting of curved space-time.

Let us now introduce the Axioms required for the construction of a covariant
Hamiltonian theory of GR, to be based on such a synchronous Lagrangian
formulation. This is achieved by imposing the following physical
requirements:

\begin{enumerate}
\item Axiom \#1: there exists a 4-scalar functional $S_{H}\left( Z,\widehat{Z%
}\right) $ defined as%
\begin{equation}
S_{H}\left( Z,\widehat{Z}\right) \equiv \int d\Omega L\left( Z,\widehat{%
\nabla }_{\mu }Z,\widehat{Z}\right) ,
\end{equation}%
where $L$ is the 4-scalar variational Lagrangian density, which carries at
most first-order derivatives $\widehat{\nabla }_{\mu }Z$ (generalized
velocities). The latter is assumed to be a quadratic dependence only. Notice
that the covariant derivative in the functional dependence of $L$ is taken
to be fixed, namely expressed by the Christoffel symbols represented in
terms of the prescribed metric tensor $\widehat{g}_{\alpha \beta }$. By
assumption, the corresponding Euler-Lagrange equations associated with the
synchronous variational principle must recover the Einstein equation for the
metric tensor $g_{\mu \nu }$ when the identification $g_{\mu \nu }=\widehat{g%
}_{\mu \nu }$ is made everywhere in the curved space-time $\mathbb{D}^{4}$.

\item Axiom \#2: the functional $S_{H}\left( Z,\widehat{Z}\right) $ is
defined in a suitable functional class of variations $\left\{ Z\right\} $.
For this purpose the ensemble field $Z$ is identified with the canonical set
$Z=x$, where%
\begin{equation}
\left\{ x\right\} =\left\{ q^{\mu \nu },p_{\mu \nu }^{\alpha }\right\} ,
\label{canonical-set}
\end{equation}%
which belongs to the 20-dimensional phase-space spanned by the state vector $%
x$. In particular, $p_{\mu \nu }^{\alpha }$ denotes the canonical momentum
conjugate to the coordinate field\ $q^{\mu \nu }$, to be defined in terms of
the Lagrangian density $L$ as%
\begin{equation}
p_{\mu \nu }^{\alpha }=\frac{\partial L}{\partial \left( \widehat{\nabla }%
_{\alpha }q^{\mu \nu }\right) }.  \label{pppp}
\end{equation}%
For completeness we introduce also the corresponding prescribed state $%
\left\{ \widehat{x}\right\} =\left\{ \widehat{q}^{\mu \nu },\widehat{p}_{\mu
\nu }^{\alpha }\right\} $ in which $\widehat{q}^{\mu \nu }$ identifies $%
\widehat{Z}$, while $\widehat{p}_{\mu \nu }^{\alpha }$ follows by evaluating
the rhs of Eq.(\ref{pppp}) for $Z=\widehat{Z}$. The functional class is
therefore identified with the synchronous canonical class%
\begin{equation}
\left\{ x\right\} \equiv \left\{
\begin{array}{c}
x(r):x(r)\in C^{2}(\mathbb{R}^{4}); \\
x\left( r\right) |_{\partial D^{4}}=x_{D}\left( r\right) \\
\delta \left( d\Omega \right) =0%
\end{array}%
\right\} .
\end{equation}%
Notice that no additional constraint is placed, in analogy with the\
functional class $\left\{ Z\right\} _{E-S}$ defined above. Here, however,
the ensemble of fields $x$ identifies a symplectic structure.

\item Axiom \#3: the continuum Hamiltonian system. The canonical fields $%
\left\{ q^{\mu \nu },p_{\mu \nu }^{\alpha }\right\} $ belonging to the
functional class $\left\{ x\right\} $ must obey the covariant Hamiltonian
equations%
\begin{eqnarray}
\frac{\partial H}{\partial p_{\mu \nu }^{\alpha }} &=&\widehat{\nabla }%
_{\alpha }q^{\mu \nu },  \label{hamil-1} \\
\frac{\partial H}{\partial q^{\mu \nu }} &=&-\widehat{\nabla }_{\alpha
}p_{\mu \nu }^{\alpha },  \label{hamil-2}
\end{eqnarray}%
where $H=H\left( q^{\mu \nu },p_{\mu \nu }^{\alpha }\right) $ is the
Hamiltonian density%
\begin{equation}
H=p_{\mu \nu }^{\alpha }\widehat{\nabla }_{\alpha }q^{\mu \nu }-L.
\end{equation}%
The set $\left\{ H,x\right\} $, with $x$ belonging to the functional class $%
\left\{ x\right\} $ and the fields satisfying Eqs.(\ref{hamil-1}) and (\ref%
{hamil-2}), prescribes a so-called continuum Hamiltonian system.

\item Axiom \#4: principle of general covariance. The manifest covariance
property of the theory must hold both for all the variational and extremal
quantities. This means that the canonical fields $\left\{ q^{\mu \nu
},p_{\mu \nu }^{\alpha }\right\} $ must be tensorial fields, so that the
variational Hamiltonian density $H$ is a 4-scalar.
\end{enumerate}

\bigskip

Let us briefly comment the physical motivations behind the Axioms. First we
notice that the Axioms generalize the approach by De Donder and Weyl to the
curved space-time. In particular, the choice of the synchronous variational
principle is a natural one corresponding to their approach in flat-space
time, because the 4-volume element is treated as an invariant in the
variational principle and at the same time is a 4-scalar. As a basic
consequence, necessarily the variational Lagrangian density is a 4-scalar.
The feature regarding the tensorial property of the canonical fields is
maintained, by replacing the partial derivatives with the covariant ones.
Finally, the previous features warrant the manifest covariance property of
the theory.

\section{The canonical theory of Einstein equation}

Based on the axiomatic formulation given above, in this section we proceed
with the formulation of a manifest covariant Hamiltonian (or canonical)%
\textbf{\ }theory for the Einstein field equations. The starting point is
the identification of the functional setting, in particular the definitions
of the canonical coordinates. In fact the latter prescribe automatically the
momenta in terms of the variational Lagrangian density. A possible
(non-unique) choice of the Lagrangian density $L$ is provided by THM.1 and
its Corollaries. In the case of the vacuum Einstein equation this is
prescribed by%
\begin{equation}
L\left( Z,\widehat{\nabla }_{\mu }Z,\widehat{Z}\right) \equiv L_{2}\left( Z,%
\widehat{\nabla }_{\mu }Z,\widehat{Z}\right) ,  \label{definition-L}
\end{equation}%
where $L_{2}\left( Z,\widehat{\nabla }_{\mu }Z,\widehat{Z}\right) $ is given
by Eq.(\ref{L4}). Instead, in the case of non-vacuum sources and possibly
non-vanishing cosmological constant the Lagrangian density must be defined as%
\begin{equation}
L\left( Z,\widehat{\nabla }_{\mu }Z,\widehat{Z}\right) \equiv L_{2}\left( Z,%
\widehat{\nabla }_{\mu }Z,\widehat{Z}\right) +L_{\Lambda }\left( Z,\widehat{Z%
}\right) +L_{1F}\left( Z,\widehat{Z}\right) ,
\end{equation}%
as given by Corollaries 1 and 2 to THM.1. It is important to remark at this
point that the form of the Lagrangian is actually gauge-dependent. In fact,
the gauge term can be identified either with an arbitrary real constant, an
exact differential, an arbitrary real function of the extremal curves or
more generally by arbitrary functions such that the same functions as well
as their variations vanish identically for the extremal fields when also the
replacement $\overline{g}_{\mu \nu }=g_{\mu \nu }$ is made.

As a consequence, by identifying $Z\equiv q^{\mu \nu }\equiv g^{\mu \nu }$,
the canonical momenta are identified with $p_{\mu \nu }^{\alpha }\equiv \Pi
_{\mu \nu }^{\alpha }$, where%
\begin{equation}
\Pi _{\mu \nu }^{\alpha }=\frac{\partial L\left( Z,\widehat{\nabla }_{\mu }Z,%
\widehat{Z}\right) }{\partial \left( \widehat{\nabla }_{\alpha }g^{\mu \nu
}\right) }=\kappa h\left( Z,\widehat{Z}\right) \widehat{\nabla }^{\alpha
}g_{\mu \nu }.  \label{pai-1}
\end{equation}%
Therefore, it follows that the canonical state can be represented as $%
\left\{ x\right\} =\left\{ Z,\Pi _{\mu \nu }^{\alpha }\right\} $. By
construction $\Pi _{\mu \nu }^{\alpha }$ is symmetric in the lower indices,
in the sense that $\Pi _{\mu \nu }^{\alpha }=\Pi _{\nu \mu }^{\alpha }$. It
must be stressed that in this context, the canonical momentum $\Pi _{\mu \nu
}^{\alpha }$ is non-vanishing since $g_{\mu \nu }\neq \overline{g}_{\mu \nu
} $ and must be considered as non-extremal. On the other hand, the extremal
value of $\Pi _{\mu \nu }^{\alpha }$, namely $\overline{\Pi }_{\mu \nu
}^{\alpha }=\kappa \overline{\nabla }^{\alpha }\overline{g}_{\mu \nu }$
vanishes identically (the same property holds also for the prescribed
fields). We notice that, thanks to the quadratic dependence required by
Axiom \#1 for the Lagrangian with respect to the generalized velocity, Eq.(%
\ref{pai-1}) in invertible. As a consequence, it recovers a form analogous
to the customary relationship between momenta and generalized velocities
occurring in relativistic particle dynamics, where the factor $\kappa
h\left( Z,\widehat{Z}\right) $ plays the role of the rest mass. This
similarity justifies the introduction of the normalization coefficient $%
\frac{1}{2}$ in the corresponding Lagrangian term.

In view of the previous considerations, the following proposition holds.

\bigskip

\textbf{THM.2 - Manifest covariant Hamiltonian theory}

\emph{Given validity of THM.1 and invoking the definition of conjugate
canonical momentum given by Eq.(\ref{pai-1}), it follows that:}

T2$_{1}$\emph{)\ The Hamiltonian density }$H=H\left( x,\widehat{x}\right) $
\emph{associated with the Lagrangian }$L\left( Z,\widehat{\nabla }_{\mu }Z,%
\widehat{Z}\right) $ \emph{is provided by the Legendre transform}%
\begin{equation}
L\left( Z,\widehat{\nabla }_{\mu }Z,\widehat{Z}\right) \equiv \Pi _{\mu \nu
}^{\alpha }\widehat{\nabla }_{\alpha }g^{\mu \nu }-H\left( x,\widehat{x}%
\right) ,  \label{legendre-hh}
\end{equation}%
\emph{where}%
\begin{equation}
L\left( Z,\widehat{\nabla }_{\mu }Z,\widehat{Z}\right) =-\kappa \left[
g^{\mu \nu }\widehat{R}_{\mu \nu }-\frac{1}{2}\widehat{\nabla }^{\alpha
}g_{\mu \nu }\widehat{\nabla }_{\alpha }g^{\mu \nu }\right] h\left( g^{\mu
\nu },\widehat{g}^{\mu \nu }\right) .  \label{L-thm2}
\end{equation}%
\emph{Here all the hatted quantities are evaluated with respect to the
prescribed metric tensor\ }$\widehat{g}^{\mu \nu }$\emph{\ and the function }%
$h\left( g^{\mu \nu },\widehat{g}^{\mu \nu }\right) $ \emph{is defined above
by Eq.(\ref{hh}). Then, written in canonical variables }$L\left( Z,\widehat{%
\nabla }_{\mu }Z,\widehat{Z}\right) =L\left( x,\widehat{x}\right) $\emph{,
it follows}%
\begin{equation}
L\left( x,\widehat{x}\right) =-\kappa hg^{\mu \nu }\widehat{R}_{\mu \nu }+%
\frac{1}{2\kappa h}\Pi _{\mu \nu }^{\alpha }\Pi _{\alpha }^{\mu \nu }.
\label{L-hat-thm2}
\end{equation}

T2$_{2}$\emph{)\ The Hamiltonian density }$H\left( x,\widehat{x}\right) $
\emph{is given by}%
\begin{equation}
H\left( x,\widehat{x}\right) =\frac{1}{2}\frac{1}{\kappa h}\Pi _{\mu \nu
}^{\alpha }\Pi _{\alpha }^{\mu \nu }+\kappa hg^{\mu \nu }\widehat{R}_{\mu
\nu }.  \label{h-finale}
\end{equation}

T2$_{3}$\emph{)\ Let us introduce the Hamiltonian action functional}%
\begin{eqnarray}
S_{H}\left( x,\widehat{x}\right) &=&\int d\Omega L\left( x,\widehat{x}\right)
\notag \\
&=&\int d\Omega \left[ \Pi _{\mu \nu }^{\alpha }\widehat{\nabla }_{\alpha
}g^{\mu \nu }-H\left( x,\widehat{x}\right) \right] ,
\end{eqnarray}%
\emph{where }$L\left( x,\widehat{x}\right) $\emph{\ is given by Eq.(\ref%
{L-hat-thm2}) and }$\left\{ x\right\} =\left\{ Z,\Pi _{\mu \nu }^{\alpha
}\right\} $\emph{\ is prescribed so that }$Z$ \emph{belongs to the
functional class }$\left\{ Z\right\} _{E-S}$ \emph{defined by Eq.(\ref{Z-E-S}%
), while }$\Pi _{\mu \nu }^{\alpha }$ \emph{belongs to the set }%
\begin{equation}
\left\{ \Pi \right\} _{E-S}\equiv \left\{
\begin{array}{c}
\Pi _{\mu \nu }^{\alpha }\left( r\right) \in C^{k}\left( \mathbb{D}%
^{4}\right) \\
\widehat{\Pi }_{\mu \nu }^{\alpha }\left( r\right) =0 \\
\Pi _{\mu \nu }^{\alpha }\left( r\right) |_{\partial \mathbb{D}^{4}}=\Pi
_{\mu \nu \mathbb{D}}^{\alpha }\left( r\right) \\
\delta \widehat{\Pi }_{\mu \nu }^{\alpha }\left( r\right) =\delta \overline{%
\Pi }_{\mu \nu }^{\alpha }\left( r\right) =0%
\end{array}%
\right\} ,
\end{equation}%
\emph{where }$\overline{\Pi }_{\mu \nu }^{\alpha }\left( r\right) $ \emph{is
the extremal curve solution of the corresponding continuum Hamilton equation
with prescribed boundary conditions. Then, the synchronous Hamiltonian
variational principle}%
\begin{equation}
\delta S_{H}\left( x,\widehat{x}\right) \equiv \left. \frac{d}{d\alpha }\Psi
(\alpha )\right\vert _{\alpha =0}=0
\end{equation}%
\emph{is required to hold for arbitrary independent variations }$\delta
g^{\mu \nu }\left( r\right) $\emph{\ and }$\delta \Pi _{\mu \nu }^{\alpha
}\left( r\right) $\emph{\ in the respective functional classes. Here }$\Psi
(\alpha )$\emph{\ is the smooth real function }$\Psi (\alpha
)=S_{H}(Z+\alpha \delta Z,\widehat{Z})$\emph{, being }$\alpha \in \left] -1,1%
\right[ $\emph{\ to be considered independent of }$g^{\mu \nu }$\emph{, }$%
\Pi _{\mu \nu }^{\alpha }$\emph{\ and }$r^{\mu }$\emph{.}\textit{\ }\emph{%
Furthermore, }$\delta g^{\mu \nu }\left( r\right) $\emph{\ and }$\delta \Pi
_{\mu \nu }^{\alpha }\left( r\right) $\emph{\ are defined respectively as in
Eq.(\ref{new-orig}) and}%
\begin{equation}
\delta \Pi _{\mu \nu }^{\alpha }\left( r\right) =\Pi _{\mu \nu }^{\alpha
}\left( r\right) -\Pi _{\mu \nu ,extr}^{\alpha }\left( r\right) ,
\end{equation}%
\emph{where here }$\Pi _{\mu \nu ,extr}^{\alpha }\left( r\right) \equiv
\overline{\Pi }_{\mu \nu }^{\alpha }\left( r\right) $.

\emph{The corresponding variational derivatives yield the continuum Hamilton
equations}%
\begin{eqnarray}
\frac{\delta S_{H}\left( x,\widehat{x}\right) }{\delta g^{\mu \nu }\left(
r\right) } &\equiv &-\frac{\partial H\left( x,\widehat{x}\right) }{\partial
g^{\mu \nu }}-\widehat{\nabla }_{\alpha }\Pi _{\mu \nu }^{\alpha }=0,
\label{ham-var-eq1} \\
\frac{\delta S_{H}\left( x,\widehat{x}\right) }{\delta \Pi _{\mu \nu
}^{\alpha }\left( r\right) } &\equiv &\widehat{\nabla }_{\alpha }g^{\mu \nu
}-\frac{\partial H\left( x,\widehat{x}\right) }{\partial \Pi _{\mu \nu
}^{\alpha }\left( r\right) }=0.  \label{ham-var-eq.2}
\end{eqnarray}%
\emph{Written explicitly, these become}%
\begin{eqnarray}
\widehat{\nabla }_{\alpha }\Pi _{\mu \nu }^{\alpha } &=&-\frac{\partial
H\left( x,\widehat{x}\right) }{\partial g^{\mu \nu }},  \label{ham-thm2-1} \\
\widehat{\nabla }_{\alpha }g^{\mu \nu } &=&\frac{1}{\kappa h}\Pi _{\alpha
}^{\mu \nu },  \label{ham-thm2-bis}
\end{eqnarray}%
\emph{so that the second one recovers as usual the definition of the
canonical momentum. These equations coincide with the Euler-Lagrange
equation (\ref{prova}) given in THM.1.}

T2$_{4}$\emph{)\ For extremal curves, provided }$\widehat{g}_{\alpha \beta }$
\emph{is identified with the extremal solution, the previous equations
reduce respectively to the Einstein equation and the condition }$\overline{%
\Pi }_{\mu \nu }^{\alpha }=0$.

T2$_{5}$\emph{)\ The Hamiltonian density }$H\left( x,\widehat{x}\right) $
\emph{given by Eq.(\ref{h-finale}) can always be replaced by the equivalent
4-scalar density}%
\begin{equation}
H_{G}\left( x,\widehat{x}\right) =H\left( x,\widehat{x}\right) +G(\widehat{x}%
,r),  \label{h-G}
\end{equation}%
\emph{where the gauge }$G(\widehat{x},r)$\emph{\ can be set so that
identically}%
\begin{equation}
H_{G}\left( \overline{x},\overline{x}\right) =0.  \label{H-G-0}
\end{equation}

\emph{Proof - }The proof of proposition T2$_{1}$) follows directly from
Axioms \#2-\#3 once the Lagrangian is identified according to Eq.(\ref%
{definition-L}), i.e. based on THM.1. Then the representation of $L\left( x,%
\widehat{x}\right) $ in terms of canonical variables in Eq.(\ref{L-hat-thm2}%
) follows by the validity of Eq.(\ref{pai-1}). Similarly, the proof of T2$%
_{2}$) follows from elementary algebra by combining Eqs.(\ref{pai-1}), (\ref%
{legendre-hh}) and (\ref{L-hat-thm2}).

To detail the proof of proposition T2$_{3}$) it is sufficient to consider
the explicit evaluation of the Euler-Lagrange equations. In particular, from
the synchronous variational principle, the variational derivative with
respect to $\delta g^{\mu \nu }\left( r\right) $ gives identically Eq.(\ref%
{ham-var-eq1}), where
\begin{equation}
\frac{\partial H\left( x,\widehat{x}\right) }{\partial g^{\mu \nu }}=\frac{1%
}{4}\frac{1}{\kappa h^{2}}\Pi _{\gamma \beta }^{\alpha }\Pi _{\alpha
}^{\gamma \beta }g_{\mu \nu }+\kappa h\widehat{R}_{\mu \nu }-\frac{1}{2}%
\kappa g^{\alpha \beta }\widehat{R}_{\alpha \beta }g_{\mu \nu }.
\end{equation}%
This provides the explicit representation for the rhs of Eq.(\ref{ham-thm2-1}%
). To prove the identity with Eq.(\ref{prova}) one has first to substitute
Eq.(\ref{ham-thm2-bis}) in the first term on the lhs of Eq.(\ref{ham-thm2-1}%
) and then evaluate explicitly the rhs of the same equation in terms of the
Hamiltonian density (to be regarded as a function of $g^{\mu \nu }$ and $%
\widehat{\nabla }_{\alpha }g^{\mu \nu }$). This gives%
\begin{equation}
\widehat{\nabla }_{i}\left[ h\widehat{\nabla }^{i}g_{\mu \nu }\right] =-%
\frac{\partial }{\partial g^{\mu \nu }}\left[ g^{\mu \nu }\widehat{R}_{\mu
\nu }h\right] +\frac{1}{2}\widehat{\nabla }^{k}g_{\mu \nu }\widehat{\nabla }%
_{k}g^{\mu \nu }\frac{\partial }{\partial g^{\mu \nu }}h,
\end{equation}%
which completes the proof of T2$_{3}$). The proof of proposition T2$_{4}$)
is achieved by noting first that, by definition $\overline{\nabla }^{\alpha }%
\overline{g}_{\mu \nu }=0$, so that Eq.(\ref{ham-thm2-bis}) implies $%
\overline{\Pi }_{\mu \nu }^{\alpha }=0$. As a consequence, the lhs of Eq.(%
\ref{ham-thm2-1}) vanishes identically, while, upon identifying $\widehat{g}%
_{\alpha \beta }$ with the extremal solution and recalling that $h\left(
\widehat{g}_{\alpha \beta }\right) =1$, the rhs gives%
\begin{equation}
\left. \frac{\partial H\left( x,\widehat{x}\right) }{\partial g^{\mu \nu }}%
\right\vert _{\overline{g}_{\mu \nu },\overline{\Pi }_{\mu \nu }^{\alpha
}}=\kappa \overline{R}_{\mu \nu }-\frac{1}{2}\kappa \overline{g}^{\alpha
\beta }\overline{R}_{\alpha \beta }\overline{g}_{\mu \nu }=0,
\end{equation}%
which coincides with the correct Einstein vacuum equation. Finally, the
proof of T2$_{5}$) follows invoking the gauge invariance property of the
corresponding Lagrangian density (see Eq.(\ref{L-gauge0})).

\textbf{Q.E.D.}

\bigskip

For completeness, we notice that the Lagrangian density $L$ given in Eq.(\ref%
{legendre-hh}) can obviously be considered as a function of the canonical
state once the exchange term (namely the first term on the rhs of the same
equation) is expressed in terms of the canonical variables.

\bigskip

Let us now consider the case of the non-vacuum Einstein equation in order to
determine the corresponding form of the Hamiltonian, required for the
validity of THM.2 in such a case. The recipe is elementary, by noting that
by assumption the source terms do not carry contributions to the canonical
momenta, as they depend only on $\widehat{g}_{\alpha \beta }$, $g_{\alpha
\beta }$ and $r$. For completeness, we include the expression due to the
possible non-vanishing cosmological constant and the source fields. Invoking
the Corollaries 1 and 2 to THM.1 given above, the resulting Hamiltonian is
found to be%
\begin{equation}
H_{tot}\left( x,\widehat{x}\right) =H\left( x,\widehat{x}\right) -L_{\Lambda
}\left( Z,\widehat{Z}\right) -L_{1F}\left( Z,\widehat{Z}\right) ,
\label{h-tot}
\end{equation}%
where $L_{\Lambda }\left( Z,\widehat{Z}\right) $ is given by Eq.(\ref%
{L-lambda}) and $L_{1F}\left( Z,\widehat{Z}\right) $ by Eq.(\ref{L1-effe}).
See also Ref.\cite{syn} for further related discussion concerning the
specific realization of the Lagrangian density $L_{1F}\left( Z,\widehat{Z}%
\right) $.

\section{Classical 4-vector Poisson brackets}

In this section we develop the formalism of Poisson brackets appropriate for
the treatment of classical fields in the context of GR according to the
DeDonder-Weyl formalism. In this respect, the peculiarity of the covariant
canonical formalism developed here must be stressed. In fact, our goal is to
reach a description of the canonical equations (\ref{ham-thm2-1}) and (\ref%
{ham-thm2-bis}) in terms of local Poisson brackets. This choice departs from
the traditional approach for continuum fields used in part of the previous
literature \cite{Goldstein}.

Let us consider first two 4-scalars $A$ and $B$ and then two arbitrary
tensor fields $\mathbf{A}\equiv A^{\alpha _{1}..\alpha _{n}}$ and $\mathbf{B}%
\equiv B_{\beta _{1}..\beta _{m}}$, generally of different order, so that $%
n\neq m$, and all to be considered smoothly dependent only on the canonical
set $\left\{ Z\right\} $ defined by Eq.(\ref{canonical-set}). Then,
respectively for $\left( A,B\right) $ and $\left( \mathbf{A},\mathbf{B}%
\right) $ the canonical Poisson brackets in terms of $\left\{ Z\right\} $
are defined in terms of the 4-vector operators%
\begin{equation}
\left[ A,B\right] _{x,j}\equiv \frac{\partial A}{\partial g^{\mu \nu }}\frac{%
\partial B}{\partial \Pi _{\mu \nu }^{j}}-\frac{\partial A}{\partial \Pi
_{\mu \nu }^{j}}\frac{\partial B}{\partial g^{\mu \nu }},  \label{aa}
\end{equation}%
\begin{equation}
\left[ \mathbf{A},\mathbf{B}\right] _{x,j}\equiv \frac{\partial \mathbf{A}}{%
\partial g^{\mu \nu }}\frac{\partial \mathbf{B}}{\partial \Pi _{\mu \nu }^{j}%
}-\frac{\partial \mathbf{A}}{\partial \Pi _{\mu \nu }^{j}}\frac{\partial
\mathbf{B}}{\partial g^{\mu \nu }},  \label{bb}
\end{equation}%
where\ in the second equation the two indices $\mu $ and $\nu $ saturate in
both terms. On the contrary, the index $j$ does not saturate, a feature
which is characteristic of the DeDonder-Weyl covariant theory, in which
coordinates and momenta have different tensorial ranks. Hence, in the first
case the rhs is a 4-vector, while in the second one the rhs is a tensor of
dimension $n+m+1$. The fundamental Poisson brackets holding for the
canonical set are therefore%
\begin{eqnarray}
\left[ g^{\beta \gamma },\Pi _{\beta \gamma }^{\alpha }\right] _{x,j}
&\equiv &\frac{\partial g^{\beta \gamma }}{\partial g^{\mu \nu }}\frac{%
\partial \Pi _{\beta \gamma }^{\alpha }}{\partial \Pi _{\mu \nu }^{j}}-\frac{%
\partial g^{\beta \gamma }}{\partial \Pi _{\mu \nu }^{j}}\frac{\partial \Pi
_{\beta \gamma }^{\alpha }}{\partial g^{\mu \nu }}=\delta _{\mu }^{\beta
}\delta _{\nu }^{\gamma }\delta _{j}^{\alpha },  \label{1q} \\
\left[ g^{\beta \gamma },g_{\beta \gamma }\right] _{x,j} &\equiv &\frac{%
\partial g^{\beta \gamma }}{\partial g^{\mu \nu }}\frac{\partial g_{\beta
\gamma }}{\partial \Pi _{\mu \nu }^{j}}-\frac{\partial g^{\beta \gamma }}{%
\partial \Pi _{\mu \nu }^{j}}\frac{\partial g_{\beta \gamma }}{\partial
g^{\mu \nu }}=0, \\
\left[ \Pi _{\alpha }^{\beta \gamma },\Pi _{\beta \gamma }^{\alpha }\right]
_{x,j} &\equiv &\frac{\partial \Pi _{\alpha }^{\beta \gamma }}{\partial
g^{\mu \nu }}\frac{\partial \Pi _{\beta \gamma }^{\alpha }}{\partial \Pi
_{\mu \nu }^{j}}-\frac{\partial \Pi _{\alpha }^{\beta \gamma }}{\partial \Pi
_{\mu \nu }^{j}}\frac{\partial \Pi _{\beta \gamma }^{\alpha }}{\partial
g^{\mu \nu }}=0.  \label{3q}
\end{eqnarray}

In terms of the fundamental brackets, the Poisson brackets with the
non-vacuum Hamiltonian density $H_{tot}\equiv H_{tot}\left( g^{\mu \nu },\Pi
_{\mu \nu }^{\alpha }\right) $ defined by Eq.(\ref{h-tot}) are%
\begin{eqnarray}
\left[ g^{\beta \gamma },H_{tot}\right] _{x,j} &\equiv &\frac{\partial
g^{\beta \gamma }}{\partial g^{\mu \nu }}\frac{\partial H_{tot}}{\partial
\Pi _{\mu \nu }^{j}}=\frac{\partial H_{tot}}{\partial \Pi _{\beta \gamma
}^{j}}, \\
\left[ \Pi _{\beta \gamma }^{\alpha },H_{tot}\right] _{x,j} &\equiv &-\frac{%
\partial \Pi _{\beta \gamma }^{\alpha }}{\partial \Pi _{\mu \nu }^{j}}\frac{%
\partial H_{tot}}{\partial g^{\mu \nu }}=-\delta _{j}^{\alpha }\frac{%
\partial H_{tot}}{\partial g^{\beta \gamma }}.
\end{eqnarray}%
As a consequence, the continuum canonical equations (\ref{ham-thm2-1})\ and (%
\ref{ham-thm2-bis}) can be extended to the non-vacuum case and equivalently
represented in terms of the 4-vector Poisson brackets as%
\begin{eqnarray}
\widehat{\nabla }_{\alpha }\Pi _{\mu \nu }^{\alpha } &=&\left[ \Pi _{\mu \nu
}^{\alpha },H_{tot}\right] _{x,\alpha },  \label{pb-1} \\
\widehat{\nabla }_{\alpha }g^{\mu \nu } &=&\left[ g^{\mu \nu },H_{tot}\right]
_{x,\alpha }.  \label{pb2-2}
\end{eqnarray}%
The fundamental Poisson brackets Eqs.(\ref{3q})-(\ref{3q}) together with the
continuum Hamilton equations (\ref{pb-1}) and (\ref{pb2-2}) display the
Hamiltonian structure characteristic of the synchronous variational
principle given in THM.2.

\section{Discussion and physical implications}

Let us now analyze the main physical aspects and the interpretation of the
theory developed here.

The first basic feature is that the boundary-value problem associated with
the Einstein equation (both in vacuum and non-vacuum) has been shown to be
characterized by an intrinsic Hamiltonian structure. Remarkably, such a
property arises when a manifestly-covariant variational approach is adopted
based on the synchronous variational principles given in THMs.1 and 2. In
particular, in this framework both the Lagrangian and the Hamiltonian
structures of the theory naturally emerge when appropriate superabundant
tensorial variables are adopted. In the case of the Lagrangian treatment,
this involves the adoption of a 10-dimensional configuration space spanned
by the symmetric variational tensor field $Z\equiv g^{\mu \nu }\left(
r\right) $. Instead, the corresponding phase-space is determined by the
Lagrangian state, namely the 50-dimensional tensor field $\left( Z,\widehat{%
\nabla }_{\mu }Z\right) $. In fact, based on the continuum Lagrangian
representation determined according to THM.1, the canonical formalism is
obtained following the DeDonder-Weyl approach, namely introducing canonical
momenta and the Hamiltonian density. This means that also the canonical
state $\left\{ x\right\} =\left\{ Z,\Pi _{\mu \nu }^{\alpha }\right\} $ is
50-dimensional.

A notable feature of the variational approach, both for the Lagrangian and
Hamiltonian treatments, is the use of a synchronous variational principle.
This is realized by introducing in the action functional a prescribed metric
tensor $\widehat{g}^{\beta \gamma }$, which remains unaffected by the
synchronous variations and is considered independent of the corresponding
variational tensor field $g^{\beta \gamma }$ as well its extremal value. The
latter is determined by the solution of the boundary-value problem of the
Euler-Lagrange equations determined by the variational principle.

Starting point is the physical interpretation of the Lagrangian variational
approach given in THM.1. In fact, the form of the modified action functional
adopted here, which differs from that given in Ref.\cite{syn}, has been
introduced specifically to permit the corresponding Hamiltonian formulation
given in THM.2. In particular, in the Lagrangian density, the new term $%
\Delta L_{1}\left( Z,\widehat{\nabla }_{\mu }Z,\widehat{Z}\right) $ given by
Eq.(\ref{delta-L1}) plays the role analogous to the kinetic energy in
particle dynamics, being quadratic in the generalized velocity $\widehat{%
\nabla }_{k}g^{\mu \nu }$. As shown by Corollary 1 to THM.1, the Einstein
equation is recovered for the extremal $g^{\mu \nu }\left( r\right) $ when
the prescribed metric tensor $\widehat{g}^{\mu \nu }\left( r\right) $ is
suitably identified. Remarkably, however, the Lagrangian structure holds for
arbitrary smooth metric tensor fields $\widehat{g}^{\mu \nu }\left( r\right)
$. In such a case $g^{\mu \nu }\left( r\right) $ obeys an Euler-Lagrange
equation which generally departs from the standard Einstein form.

Let us now consider the Hamiltonian formulation given by THM.2. Basic
feature of the treatment is that the Lagrangian density is regular, in
analogy with the Lagrangian treatment well-known in particle dynamics.
Indeed, the transformation from the Lagrangian state to the corresponding
Hamiltonian one is non-singular, being realized by a smooth bijection. As
for the Lagrangian treatment, also the Hamiltonian structure emerges only
when tensorial superabundant variables are adopted in the framework of the
synchronous principle. The feature is essential because it permits to
display the contribution of the canonical momenta when they are
non-vanishing. This property occurs when $g_{\mu \nu }\neq \widehat{g}_{\mu
\nu }$, namely the variational curves differ from the prescribed metric
tensor. In fact, from the definition (\ref{pai-1}), in such a case one has
that $\Pi _{\mu \nu }^{\alpha }\neq 0$, since generally $\widehat{\nabla }%
^{\alpha }g_{\mu \nu }\neq 0$. This feature turns out to be instrumental for
the construction of the continuum canonical equations given in THM.2 as well
as the subsequent formulation in terms of Poisson brackets.

On the other hand, once the constraint $g_{\mu \nu }\equiv \widehat{g}_{\mu
\nu }$ is set on the extremal equations, the Hamiltonian structure $\left\{
H_{tot},x\right\} $ is only apparently lost, because of the vanishing of the
canonical momenta in such a case. In fact, in this limit the state $x$
collapses\ to the 10-dimensional hypersurface of the phase-space spanned by
the canonical state $x=\left[ g_{\mu \nu },0\right] \equiv x_{0}$ having
identically-vanishing momenta. Nevertheless, inspection of the Hamilton
equations shows that they actually hold also when $x=x_{0}$ is set, namely
the partial derivatives of the Hamiltonian density are evaluated at such
state. Hence, the Hamiltonian structure indeed is preserved also in this
case. This conclusion is of fundamental importance, for its physical
implication. In fact it identifies uniquely the covariant continuum
Hamiltonian system associated with the Einstein equation. We remark that
this result is ultimately due to the validity of the synchronous variational
principle formulated in THM.2, which is based on the introduction of the
constrained functional class . Such a feature is not unique in classical
physics. In fact, in this regard, one may view the metric-compatibility
condition $\overline{\nabla }_{\alpha }\overline{g}^{\mu \nu }=0$ as playing
a role analogous to the mass-shell condition $u^{\mu }u_{\mu }=1$ in the
so-called super-Hamiltonian approaches to classical particle dynamics. In
both cases, the constraints do not affect the validity of the corresponding
Hamiltonian structure, since they hold only for the extremal curves.

It must be remarked that the adoption of the superabundant field variables $%
\left\{ g_{\mu \nu },\widehat{g}_{\mu \nu }\right\} $ permits one to avoid
the introduction of extra-dimensions in the space-time. The physical
interpretation of such a representation is as follows: 1) $g_{\mu \nu }$ is
a \emph{tensor field}, whose dynamical equations are provided by the
Euler-Lagrange equations determined by the synchronous variational
principle. In this sense, in the action functional $g_{\mu \nu }$ has
initially no geometrical interpretation, since it does not raise or lower
indices nor it appears in the covariant derivatives or the Ricci tensor $%
\widehat{R}_{\mu \nu }$. 2)\ Instead, $\widehat{g}_{\mu \nu }$ is the \emph{%
geometrical field}. In fact, it determines a number of relevant geometric
properties: the invariant 4-volume element, the covariant derivatives and
the Ricci tensor $\widehat{R}_{\mu \nu }$ and finally it raises/lowers
tensor indices. 3) The contribution $g^{\mu \nu }\widehat{R}_{\mu \nu }$
appearing in both the Lagrangian and Hamiltonian densities can be
interpreted as an effective coupling term between the physical field $g^{\mu
\nu }$ and the geometrical quantity $\widehat{R}_{\mu \nu }$. In this sense,
this term is similar to the well-known coupling term $A_{\mu }J^{\mu }$
occurring in the corresponding Lagrangian formulation for the EM field.

Finally, we notice that there is another possible interpretation of the
formalism developed here, and in particular the distinction between the
metric tensors $g_{\mu \nu }$ and $\widehat{g}_{\mu \nu }$. This is provided
by the so-called \textit{induced gravity} (or \textit{emergent gravity}),
namely the idea that the space-time geometrical properties emerge as a mean
field approximation of an underlying microscopic stochastic or quantum
degrees of freedom. Indeed, it is tempting to view $\widehat{g}_{\mu \nu }$
as a macroscopic mean field emerging on a (fluctuating) background
represented by the variational field $g_{\mu \nu }$. Such an idea is
naturally connected with the process of quantization. As shown in the
present paper, this gives rise to a new type of action which depends both on
$g_{\mu \nu }$ and $\widehat{g}_{\mu \nu }$ and permits the identification
of a covariant Hamiltonian structure associated with the classical
gravitational field. Remarkably, the customary non-vacuum Einstein equation
\textit{emerges} when the Hamiltonian structure is collapsed on the
10-dimensional subset of phase-space on which canonical momenta vanish
identically.

\section{Conclusions}

Historically, the problem of the identification of the manifestly covariant
Hamiltonian structure associated with the Einstein equation of General
Relativity (both in vacuum and non-vacuum) and which retains also the
correct gauge transformation properties, has remained apparently unsolved to
date. Actual reasons remain substantially a guesswork. For example one such
possible conjecture lies in the strong influence set by some of the
most-distinguished authors dealing with the subject. One example among them
can probably be ascribed to Dirac himself. This refers, in particular, to
his famous 1948 treatment of relativistic Hamiltonian systems based on
constrained dynamics. In his approach, in fact, although overall covariance
of the Hamilton equations is still warranted, explicit non-tensor
Hamiltonian variables were adopted. On the other hand, there is good
evidence that Dirac approach had actually a deep influence in the subsequent
literature. This might possibly explain how and why the opposite view took
stand, namely the concept that only manifestly non-covariant Hamiltonian
treatments of GR are actually possible.

Contrary to such a viewpoint, in this paper the route of the manifestly
covariant Hamiltonian theory has been pursued to a full extent, enabling us
to establish a manifestly covariant Hamiltonian formulation for Einstein
equation. The goal has been reached by generalizing the so-called
\textquotedblleft DeDonder-Weyl\textquotedblright\ formalism in the context
of curved space-time and the introduction of a suitable kind of synchronous
Hamiltonian variational principles extending analogous Lagrangian
formulations earlier recently developed in the literature.

Several remarkable new features emerge.

The first one is the adoption of a constrained variational formulation, the
constraints being intrinsically related to the notion of synchronous
variation and of synchronous variational principle.

The second is the manifest covariant property of the theory at all levels,
i.e., \ beginning from the definition of the action functional itself, its
Lagrangian density, the prescription of the Hamiltonian variables as well
as, finally, the construction of the Hamiltonian Euler-Lagrange equations.

The third one involves the adoption of a superabundant canonical-variable
approach. In this reference, a critical element turns out to be the adoption
of a synchronous variational principle. This feature, in fact, permits one
to determine the Einstein equation via a Lagrangian formulation expressed in
superabundant variables. As a consequence, it has been shown that the
Lagrangian equation can be cast also in the equivalent manifestly-covariant
Hamiltonian form and in terms of continuum Poisson brackets.

The\ fundamental key feature which arises is that the Hamiltonian structure
of the Einstein equations emerges in a natural way when the manifestly
covariant approach is realized.

These features suggest the theory presented here as an extremely promising
and innovative research topic. The theory presented here is in fact
susceptible of a plethora of potential applications, besides its natural
framework, i.e., GR.\textbf{\ }In particular, possible subsequent
developments range from the investigation of the Hamiltonian and
Hamilton-Jacobi structure of GR, to cosmology as well as relativistic
quantum field theory and quantum gravity.

\bigskip

\textbf{Acknowledgments -} Work developed within the research projects of
the Czech Science Foundation GA\v{C}R grant No. 14-07753P (C.C.)\ and Albert
Einstein Center for Gravitation and Astrophysics, Czech Science Foundation
No. 14-37086G (M.T.).

\end{document}